\documentstyle[prb,aps,multicol,psfig]{revtex}
\begin{document}
\title{Phase transition in a chain of quantum vortices}
\author{C. Bruder$^{1,}$\cite{Basel}, L.~I. Glazman$^{2,3}$, 
A.~I. Larkin$^{2,4}$, J.~E. Mooij$^{3}$, and A. van Oudenaarden$^{3}$} 
\address{$^{1}$Institut f\"ur Theoretische
Festk\"orperphysik, Universit\"at Karlsruhe, D-76128 Karlsruhe,
Germany\\ $^{2}$Theoretical Physics Institute, University of
Minnesota, Minneapolis MN 55455\\ $^{3}$Delft University of
Technology, 2600 GA Delft, The Netherlands\\
$^{4}$L.~D. Landau Institute for Theoretical Physics, Moscow 117940, 
Russia}
\date{\today}
\maketitle

\begin{abstract}
We consider interacting vortices in a quasi-one-dimensional array of
Josephson junctions with small capacitance. If the charging energy of
a junction is of the order of the Josephson energy, the fluctuations
of the superconducting order parameter in the system are considerable,
and the vortices behave as quantum particles. Their density may be
tuned by an external magnetic field, and therefore one can control the
commensurability of the one-dimensional vortex lattice with the
lattice of Josephson junctions. We show that the interplay between the
quantum nature of a vortex, and the long-range interaction between the
vortices leads to the existence of a specific
commensurate-incommensurate transition in a one-dimensional vortex
lattice. In the commensurate phase an elementary excitation is a
soliton, with energy separated from the ground state by a finite
gap. This gap vanishes in the incommensurate phase. Each soliton
carries a fraction of a flux quantum; the propagation of solitons
leads to a finite resistance of the array. We find the dependence of
the resistance activation energy on the magnetic field and parameters
of the Josephson array. This energy consists of the above-mentioned
gap, and also of a boundary pinning term, which is different in the
commensurate and incommensurate phases. The developed theory allows us
to explain quantitatively the available experimental data.
\end{abstract}


\begin{multicols}{2}

\section{Introduction}
The interest in Josephson junction arrays in the last decade was
to a large degree prompted by the fact that these systems are suitable as a
testing ground for various predictions of quantum many-body theory
(for an overview see, e.g., Refs.~\onlinecite{Simanek94,FazioS}). If
the charging energy of a junction $E_C=e^2/2C$ is comparable with its
Josephson energy $E_J$, the phase of the superconducting order
parameter is subject to quantum fluctuations (here $C$ is the
capacitance of a junction). At some critical value of $E_C/E_J$ the
global phase coherence is destroyed, and the array becomes an
insulator\cite{Geerligs}. This transition apparently is driven by
proliferation of spontaneously created vortices, i.e., topological
excitations of the array, in which the phase of the order parameter
varies by $2\pi$ on going around a plaquette. At smaller ratios
$E_C/E_J$ vortices induced by an external magnetic field still possess
quantum properties. The vortex dynamics is particularly sensitive to
the quantum fluctuations of the phase: the vortex mass, for example,
is finite entirely due to these fluctuations \cite{LarkinOS,EckernS}.

A single vortex in a Josephson junction array behaves as a
ballistically propagating quantum particle\cite{Elion}. These
particles are strongly interacting, however: for the values of
$E_C/E_J\lesssim 1$ at which the global phase coherence is preserved,
their interaction energy $U_{v-v}\propto E_J$ is larger than the 
bandwidth for a single vortex. A finite magnetic field applied
perpendicularly to the array, creates a lattice with a vortex density
proportional to the field strength. Depending on the magnetic field
flux per plaquette, the vortex lattice is commensurate or
incommensurate with the junction array\cite{Delftconf}. The
commensurability effect exists of course even for classical vortices
in an array with $E_C/E_J\to 0$. The array acts like a periodic
potential with an amplitude $U_p\sim 0.2 E_J$ and some period $a$ (the
period of the Josephson array) for each vortex\cite{Lobb}. In a
classical system, this is expected to be a source of strong pinning, as
$U_p$ and $U_{v-v}$ are of the same order. Quantum fluctuations bring
new physics into the problem. The period of the pinning potential is
relatively small, and therefore its amplitude is suppressed readily by
quantum fluctuations. On the contrary, depending on the magnetic
field, the vortex lattice period may be significantly larger than $a$,
thus making the vortex lattice robust against quantum fluctuations.

\begin{figure}[h]
\begin{center}
\leavevmode \psfig{figure=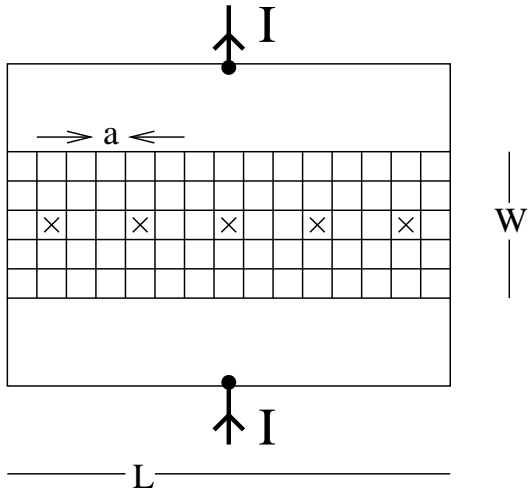,width=7cm}
\narrowtext
\caption{Quasi-one-dimensional Josephson array. Each side of a
plaquette corresponds to a single Josephson junction. Crosses denote
vortices located in the central row of the array. The properties of
the vortex chain are probed by passing a current $I$ from one
superconducting contact to the other and measuring the voltage between
them. The array width is $W$, and the size of a single plaquette is
$a\times a$.}
\label{array}
\end{center}
\end{figure}

Commensurability effects in a chain of quantum vortices were
investigated in the recent experiments of Oudenaarden {\it et
al.}\cite{Oudenaarden1}. There a number of two-dimensional arrays with
various ratios $E_C/E_J$, and various widths of the order of $10$
cells were studied. All arrays were quasi-one-dimensional in fact,
their length varying between $100$ and $1000$ cells. Superconducting
contacts parallel to the long sides of the array were providing a
potential confining the vortices to the central row of the array, see
Fig.~\ref{array}. Applying a current through the leads (perpendicular
to the one-dimensional vortex chain) and measuring the resistance of
the system as a function of the magnetic field, van Oudenaarden {\it
et al.} found almost zero resistance in the regions centered around
the commensurate values of the one-dimensional vortex density. This
was interpreted as an indication of a finite-gap state (``Mott
insulator'' phase), induced by the vortex-vortex interaction in the
presence of a periodic potential. If the magnetic field was tuned away
from these special regions, a transition to a resistive state was
observed, indicating moving vortices (``conducting'' phase).

The analogy between the observed transition in the system of vortices
and the textbook Mott transition in electron systems can be made more
explicit: $E_C/E_J$ here plays the role of the ratio $t/U$ of the
electron bandwidth to the on-site repulsion potential; the magnetic
field $B$, creating vortices, plays the role of the electron chemical
potential $\mu$. The phase diagram in the variables $(t/U,\mu)$
consists of two phases. The insulating phase occurs at relatively
small values of $t/U\lesssim 1$. In this phase, the electron density
is constant as a function of $\mu$, and fixed by the commensurability
condition (one electron per lattice site). One can assign the value
$\mu=0$ to the line of particle-hole symmetry in this phase
diagram. Deviation from this symmetry line makes the excitation gap in
the insulating phase smaller; the gap width is a non-analytical
function of $\mu$, reflecting the violation of the particle-hole
symmetry at $\mu\neq 0$. At a certain critical value of $|\mu|$, which
depends on $t/U$, the gap disappears and a transition to a
``conducting'' phase occurs. This conventional picture is modified
somewhat in the one-dimensional case, but qualitatively remains
valid. A similar description applies to Mott transitions in Bose
systems with repulsion\cite{FisherFisher}.

The line of particle-hole symmetry in the case of vortices corresponds
to a special value $B_0$ of the magnetic field, which induces a vortex
lattice commensurate with the period of the junction array. The
analogy to the electronic case described in the last paragraph
suggests that the vortex density remains constant in a finite interval
of $B$ around $B_0$. One also expects a gap in the excitation spectrum
of the array that diminishes as a function of $|B-B_0|$ within the
commensurate phase (which is the analog of the insulating
phase). Indeed, in the experiment a cusp-like dependence of the
resistance activation energy on $|B-B_0|$ was observed (see
Ref.~\onlinecite{Oudenaarden1} and Section\ref{experiment} of this
paper). At some critical value of $|B-B_0|$ the observed temperature
dependence of the resistance becomes considerably weaker. This may
indicate that a transition to a gapless (``conducting'') phase occurs,
accompanied by the creation of discommensuration solitons.

In this paper, we present a quantitative theory of the
commensurate-incommensurate transition for a chain of quantum vortices
in a quasi-one-dimensional Josephson array (see
Ref.~\onlinecite{Kardar} for another type of
commensurate-incommensurate transition in Josephson arrays). The
transition to the incommensurate state occurs by proliferation of the
discommensuration solitons through the vortex chain. We are able to
develop a theory by analytical means, because of a special feature of
the system we consider. It turns out, that the range of the
interaction between the vortices is much longer than the inter-vortex
distance. Therefore, the solitons consist of many vortices, and
possess a large effective mass. Thus the theory for the
commensurate-incommensurate transition is essentially
classical. However, to relate the parameters of this theory to the
generic properties ($E_C$ and $E_J$) of the Josephson array, we need
to consider a single vortex as a quantum particle: the amplitude of
the periodic pinning potential depends on the bandwidth of the
vortex. After that, we are able to find analytically the excitation
gap existing in the commensurate phase and the boundaries of this
phase in the $(B, E_C/E_J)$ plane.

We relate the characteristics of the commensurate and incommensurate
phases to an observable quantity, the activation energy of the
resistance $E_R$. In the commensurate phase, the transfer of one flux
quantum between the edges of the array occurs via a sequence of
solitons propagating through it. The number of solitons necessary to
transfer one vortex is equal to the ratio of the periods of the vortex
lattice and the junction array; typically this ratio is large. At any
time during the vortex transfer, there is no more than one soliton
present in the chain. We demonstrate that $E_R$ depends not only on
the properties of the ``bulk'' one-dimensional system, but also
reflects boundary pinning effects, accompanying the passage of
vortices through the ends of the array. One soliton changes the length
of the vortex chain only by one period of the junction array, which is
less than the inter-vortex spacing. Hence, in the commensurate phase,
the process of vortex flow through the array can be viewed as motion
of a rigid vortex chain. Because of the rigidity, the vortex chain
cannot adjust itself to the boundary pinning potential. The potentials
produced by the two ends of the array add to $E_R$: the relative phase
of these two contributions depends on whether the total flux piercing
the junction array equals an integer number of flux quanta. Thus, in
the commensurate state, there are two major terms in $E_R$. The first
term is the activation energy of a soliton, and the second term is the
sum of the boundary pinning energies. This second (smaller) term
oscillates with the magnetic flux piercing the array. In the
incommensurate state, the vortex chain is compressible, and can adjust
to the boundaries of the array, if the latter is sufficiently long. As
a result, the main term in $E_R$ is the boundary pinning potential,
which a vortex has to overcome to enter the array. A correction is
provided by the finite compression energy of the chain. Its average
value depends on the compressibility of the chain, renormalized by the
solitons, and is inversely proportional to the length of the
array. This term oscillates with the flux threading the system.

The paper is organized as follows. In Section \ref{rigid}, we
introduce a model of classical vortices in the quasi-one-dimensional
Josephson array. Here we neglect the discreteness of the array, and
the screening of the vortex-vortex interaction. This approximation
means that the the vortex chain is entirely incompressible. We
establish the stability criterion for a one-dimensional vortex chain
against formation of a zigzag structure. We calculate the equilibrium
number of vortices as a function of the magnetic field and determine
the boundary pinning caused by the interaction of the vortices with
the ends of the array. For an incompressible chain, this gives us the
equilibrium position for each vortex. In the following part, Section
\ref{effpinning}, we discuss bulk pinning by reintroducing the
discreteness of the junction array. The array creates a periodic
potential for each vortex, which behaves as a quantum particle in this
potential. We demonstrate that typically the amplitude of the quantum
fluctuations of a vortex exceeds the period of the array. (This
justifies, in fact, the approximations made in Section
\ref{rigid}). We calculate the residual pinning potential, suppressed
by quantum and thermal fluctuations, acting on a single vortex.

The results of the Sections \ref{rigid} and \ref{effpinning} are
directly applicable to short arrays, i.e., arrays that are shorter
than the range of the vortex-vortex interaction. The main goal of
these sections though is to provide us with the coefficients necessary
to write down the effective Hamiltonian describing a compressible
chain in a long array. We start the next Section \ref{compressible}
with an estimate of the vortex-vortex interaction range. It is defined
by two mechanisms: (1) the effect of the magnetic field induced by the
vortices, and (2) the interaction of the vortices in the Josephson
array with the Abrikosov lattice in the contacts to the array. The
estimate demonstrates that for the conditions of the
experiments\cite{Oudenaarden1} the range indeed exceeds greatly the
inter-vortex distance, but still may be smaller than the system
length, making it necessary to account for a finite compressibility of
the vortex chain. We therefore derive the long-wavelength theory for
the compressible vortex chain. This theory enables us to describe, in
Section \ref{phases}, the commensurate-incommensurate transition. We
determine the boundaries of the commensurate phase, and find the
dependence of the activation energy for elementary excitations on the
parameters of the system. Also in this section, we discuss the
behavior of the resistivity following from the picture we
developed. We compare our results with the existing experiment in
Section \ref{experiment}. Using the experimental values of the vortex
density at the commensurate-incommensurate transition and the maximum
of the activation energy of the resistance, we are able to give
parameter-free estimates of the range of the vortex-vortex interaction
and of the elastic constant of the vortex chain. The effective pinning
potential turns out to be at least an order of magnitude smaller than
the bare potential due to quantum fluctuations (as calculated in
Section~\ref{effpinning}). The soliton length is extremely large and
of the order of the length of the array. The long-range nature of the
vortex-vortex interaction leads to a large value of the elastic
constant: the chain is virtually rigid in the incommensurate phase.
Our theory explains consistently the main experimental observations
reported in in this paper and in Ref.~\onlinecite{Oudenaarden1}: (1)
the cusp-like dependence of the activation energy on the magnetic
field in the commensurate phase; (2) the large value of this
activation energy (compared to $E_J$ and $E_C$), and (3) oscillations
of the resistance with the applied magnetic field, with a period
corresponding to one flux quantum through the entire array, in the
incommensurate phase. We conclude with a discussion in Section
\ref{discussion}.

\section{Rigid vortex chain}
\label{rigid}

We consider a two-dimensional Josephson array of lattice constant $a$,
length $L$ and width $W$ where $L\gg W$, see Fig.~\ref{array}. The
``sites'' of this array are superconducting islands, linked by
Josephson junctions that are characterized by a capacitance $C$ and a
critical current $I_C$. The phases $\varphi_{\bf i}$ of the order
parameter of the islands (numbered by vectors $\bf i$) are the only
dynamical degrees of freedom of the system. For an infinite
two-dimensional system, the Lagrangian can be written in the
standard\cite{LarkinOS} way,
\begin{equation}
L=\sum_{\langle {\bf i}, {\bf j}\rangle}\left\{{\hbar^2\over
8E_C}\left({\partial\varphi_{{\bf i}, {\bf j}}\over\partial
t}\right)^2- E_J\left[1-\cos\left(\varphi_{{\bf i}, {\bf
j}}\right)\right]\right\} \; .
\label{lagrangian}
\end{equation}
Here the sum is taken over the nearest neighbors, and $\varphi_{{\bf
i}, {\bf j}}$ is the phase difference across a link of the array,
$E_J=I_C\Phi_0/2\pi$ and $E_C=e^2/2C$ are the Josephson and charging
energy, respectively; $\Phi_0=hc/2e$ is the flux quantum. The
Lagrangian (\ref{lagrangian}) describes quantum fluctuations of the
phase in the array. At a certain critical value\cite{critical} of the
ratio $E_C/E_J\sim 1$, the proliferation of spontaneous vortices and
antivortices through the system destroys the long-range order. We
consider smaller values of $E_C/E_J$, and neglect the existence of
spontaneous topological excitations. Vortices in the spatial
distribution of the phase $\varphi$ are then induced only by an
external magnetic field $B$. A vortex is characterized by a phase
change of $2\pi$ on going around a plaquette. The effective Lagrangian
in terms of the vortex positions,
\begin{equation}
L=\sum_{i}\frac{M}{2}\left(\frac {d{\bf r}_i}{dt}\right)^2 -
\sum_{i,j}\frac{1}{2}U_{v-v}({\bf r}_i, {\bf r}_j)- \sum_{i}U_p({\bf
r}_i)
\label{lagrangian1}
\end{equation}
can be derived\cite{LarkinOS} from Eq.~(\ref{lagrangian}). Here
$M=\pi^2\hbar^2/4a^2E_C$ is the vortex mass, $U_{v-v}({\bf r}_i, {\bf
r}_j)$ is the interaction energy between the vortices, and $U_p({\bf
r}_i)$ is the pinning potential which represents the effect of a
discrete lattice of junctions on the vortex motion. In an infinite
array, the energy $U_{v-v}$ depends only on the distance between
vortices, and can be approximated by the standard expressions valid
for vortices induced in a thin superconducting
film\cite{DeGennes}. For a geometrically restricted array,
Fig.~\ref{array}, the form of the interaction potential $U_{v-v}$
depends crucially on the boundary conditions for the phase that are
set by the massive superconducting contacts. The superfluid density in
these superconducting strips exceeds greatly the effective superfluid
density in the array. Therefore, each vortex in the array is repelled
from the boundaries (this is represented by the terms $U_{v-v}({\bf
r}_i,{\bf r}_i)\equiv U_{v-v}(y_i)$ in the Lagrangian). At a
sufficiently weak magnetic field, $B\lesssim \Phi_0/W^2$, the
inter-vortex distance is large enough, and vortices occupy only the
central row of the array.

The pinning potential in this one-dimensional case may be
modeled\cite{Lobb} by a function of $x$ only,
\begin{equation}
U_p(x)=0.1E_J[1-\cos(2\pi x/a)]\; .
\label{barepotential}
\end{equation}
We will see in the next section, that quantum fluctuations of the
vortex positions strongly diminish the role of pinning by the periodic
lattice of Josephson junctions. For now, we will ignore the
contribution of the pinning potential, given by the third term in the
Lagrangian Eq.~(\ref{lagrangian1}).

Because of the large superfluid density in the contacts, the phase of
the order parameter varies only slightly along each of the long
boundaries of the array. Currents induced by a vortex in the array
flow through it almost perpendicularly to the boundaries. In the limit
of infinite superfluid density and infinite London-Pearl penetration
depth\cite{DeGennes} in the contacts, the currents within the array do
not decay with the distance from a vortex. As a result, the range of
the vortex-vortex interaction is infinite. The interaction potential,
up to an arbitrary constant, has the following form
\begin{equation}
U_{v-v}= -2\pi^2E_J|x_i-x_j|/W
\label{vortex-vortex}
\end{equation}
at $|x_i-x_j|\gtrsim W$. Because of this form, the vortex chain is
absolutely rigid at small wave vectors.

\begin{figure}[h]
\begin{center}
\leavevmode \psfig{figure=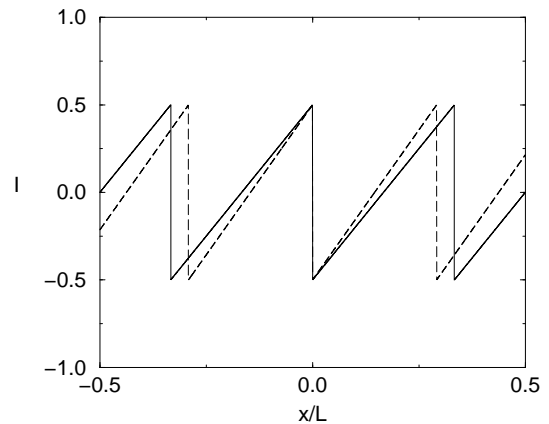,width=7cm}
\narrowtext
\caption{Current (in arbitrary units) across the array as a function
of the coordinate x along the array. Note the jumps at the vortex
positions $x_i$. For $nL=\text{integer}$ (solid line), we can identify
the two edges at $\pm L/2$, and shifting the vortex chain along the
x-direction does not change the energy, Eq.~(\protect\ref{energy}). In
contrast to that, for $nL\ne\text{integer}$ (dashed line), the energy
depends on the position of the chain (boundary pinning).}
\label{saw}
\end{center}
\end{figure}

If the smallest inter-vortex distance exceeds the array width, one can
use the limiting form of the potential (\ref{vortex-vortex}) to
calculate the contribution of the vortex-vortex interaction [the
second term in Eq.~(\ref{lagrangian1})] to the energy of the vortex
chain. It is more convenient, however, to write down this energy
directly in terms of the phase distribution in the array:
\begin{equation}
{\partial\varphi\over\partial y}(x)={1\over W}[2\pi nx-\pi\sum_{i=1}^N
\text{sign}(x-x_i) + \varphi_0]\; .
\label{gradient}
\end{equation}
Here, we have replaced the phases $\varphi_{\bf i}$ of the islands by
a continuous variable $\varphi (\bf r)$. The form (\ref{gradient}) of
the phase gradient is valid at distances $|x-x_i|\gtrsim W$ away from
the vortex centers $x_i$. The magnetic field enters via the
one-dimensional density $n=BW/\Phi_0$. The phase $\varphi_0$ has the
meaning of the average phase difference between the contacts, and will
be used as a Lagrange multiplier to enforce the condition of fixed
current $I$ in the y-direction through the array. The typical shape of
the phase gradient is illustrated in Fig.~\ref{saw}. The energy in the
presence of a current $I$ between the contacts can be written as
\begin{equation}
E(\{x_i\},\varphi_0)={E_JW\over 2} \int_{-L/2}^{L/2}dx \left(
{\partial\varphi\over\partial y}\right )^2 - {\Phi_0\over
2\pi}I\varphi_0\; .
\label{energy}
\end{equation}

The equilibrium positions of the vortices and $\varphi_0$ for a given
value of the current $I$ are defined by the set of conditions
\begin{eqnarray}
{\partial E(\{x_i\},\varphi_0)\over\partial x_i}&=0\nonumber\\
{\partial E(\{x_i\},\varphi_0)\over\partial \varphi_0}&=0\; .
\label{derivative}
\end{eqnarray}

At first we will consider the case $I=0$ and will determine the
equilibrium number $N$ and positions $x_i^0$, $i=1, ..., N$ of
vortices in the array. Solving Eqs.~(\ref{derivative}), we obtain
\cite{footnote}
\begin{equation}
N=\text{Int}(nL)\; ,
\label{eqnumber}
\end{equation}
where $\text{Int}(x)$ is the integer part of $x$. We will consider
only positive values of the magnetic field, $n>0$. The equilibrium
positions are given by
\begin{equation}
x_i^0={2i-1-N\over 2n}\; ,
\label{eqpositions}
\end{equation}
which means that the vortices are equidistant,
$x_{i+1}^0-x_i^0=1/n$. The first and the last vortex of the chain are
located at a distance $[nL-\text{Int}(nL)+1]/(2n)\ge 1/(2n)$ away from
the ends of the array. On increasing the flux, they move towards the
center.

For deviations of the vortex coordinates from their equilibrium
positions, the energy Eq.~(\ref{energy}) may be expressed as
\begin{equation}
E=E_0+{2\pi^2E_Jn\over W}\{\bar x^2\text{Frac}(nL) +\sum_{i=1}^N
(x_i-x_i^0)^2\}\; ,
\label{deviations}
\end{equation}
where $\bar x=\sum_i x_i /N$ is the center-of-mass of the vortex
chain; for brevity, hereafter we use the notation
\begin{equation}
\text{Frac}(nL)\equiv nL-\text{Int}(nL)\; .
\label{frac}
\end{equation}
Each individual vortex resides in a parabolic well, centered at the
vortex equilibrium position; this is the result of the infinite-range
interaction between the vortices.

The term proportional to $\bar x^2$ is caused by the interaction of
the vortices with the two boundaries at $\pm L/2$: if $nL$ is integer,
shifting the vortex chain along the x-direction does not change the
energy, Eq.~(\protect\ref{energy}), see Fig.~\ref{saw}. For general
values of $nL$, the energy depends on the position of the chain. That
means that the boundaries pin the vortex chain.

\begin{figure}[h]
\begin{center}
\leavevmode \psfig{figure=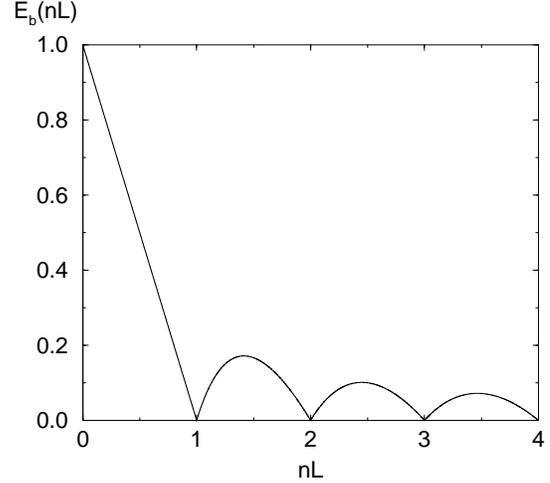,width=7cm}
\narrowtext
\caption{Activation energy (in units of $\pi^2E_JL/ 2W$) of the rigid
chain as a function of $nL=BWL/\Phi_0$. The critical current is given
by the same curve: $I_C^{\text{array}}=4E_{\text{b}}(nL)/\Phi_0$, see
Eq.~(\protect\ref{critcurrent}) and also
Ref.~\protect\onlinecite{Likharev}.}
\label{activationfigure}
\end{center}
\end{figure}

The activation energy of the system is given by the difference in
ground-state energies $E_0$ of the $N+1$ and $N$-vortex chains at a
given value of the flux density $n$. A straightforward calculation
starting with Eq.~(\ref{energy}) at $I=0$ yields
\begin{equation}
E_{\text{b}}(nL)={\pi^2E_J\over 2W
n}[1-\text{Frac}(nL)]\text{Frac}(nL)\; .
\label{activation}
\end{equation}
The boundary pinning energy (\ref{activation}) vanishes for integer
values of $nL$; the maxima between two zeroes are $\pi^2E_J/(8nW)$,
and decay with the magnetic field as $1/n$, see
Fig.~\ref{activationfigure}.

It is also possible to calculate the critical current of the array,
$I_C^{\text{array}}$. We define it as the current at which the
stability of the center-of-mass $\bar x$ is lost. The relation between
$I$ and $\bar x$ can be found from Eq.~(\ref{energy}),
\begin{eqnarray}
I={2\pi E_J\over\Phi_0}\int_{-L/2}^{L/2}dx \left ({\partial
\varphi\over\partial y} \right )\nonumber\\
={4\pi^2E_J\over\Phi_0}{\bar x\over W}\text{Frac}(nL) \; .
\label{current}
\end{eqnarray}
The center-of-mass stability requires that no vortex is to enter or
leave the system; this restriction leads to
\begin{equation}
I_C^{\text{array}}\!\! =\frac{4E_{\text{b}}}{\Phi_0}\! \equiv\! {\pi
I_C\over nW}[1-\text{Frac}(nL)]\text{Frac}(nL)\, .
\label{critcurrent}
\end{equation}
This functional dependence of the critical current on $nL$ for the
array coincides with the one obtained in Ref.~\onlinecite{Likharev}
for a thin-film bridge.

It is clear that the one-dimensional approximation breaks down for
large magnetic fields, i.e., if the vortex-vortex distance becomes
considerably less than the width $W$ of the array. To obtain a
quantitative value of the critical field, we studied the instability
of the vortex chain towards formation of a zigzag deformation. For
that purpose, we derive the full formula for $U_{v-v}({\bf r}_i, {\bf
r}_j)$, which requires properly taking into account the infinite
number of image vortices necessary to fulfill the boundary conditions
at the superconducting contacts. After that, we replace
Eq.~(\ref{energy}) by its two-dimensional analogue, which depends on
the two-dimensional vectors of displacements of each vortex. An
analysis of the dependence of this energy on the transverse vortex
displacements yields the value of the critical field at which the
zigzag pattern forms:
\begin{equation}
B_{\text{crit}}={\Phi_0 \over 0.65 W^2}\; .
\label{zigzag}
\end{equation}
In other words, the vortex distance has to be larger than $0.65 W$ for
the one-dimensional approximation to be valid.

\section{Pinning by the periodic potential}
\label{effpinning}

In the last section, the discreteness of the system was neglected
completely. At first sight, this seems to be an unreasonable
approximation. Indeed, only at $nW>(0.4/\pi^2)(W/a)^2$ the amplitude
$0.2E_J$ of the pinning potential (\ref{barepotential}) is smaller
than the variation $\delta E$ of the energy (\ref{deviations}) if a
single vortex is displaced by $a/2$ ($a/2$ is the distance between the
minimum and maximum of the pinning potential). Note that for the
stability of a single-row vortex chain, the condition $nW<1.62$ must
be satisfied (see Eq.~(\ref{zigzag})). The two restrictions on $nW$
are incompatible except for quite narrow arrays, $W/a\leq 6$. In this
section, we demonstrate that the effective pinning potential is
reduced significantly by quantum fluctuations of the vortex
coordinates, which makes the above restriction irrelevant, even at
relatively small ratios $E_C/E_J$.

Forgetting the interactions with the other vortices for a moment, each
vortex is described by a Hamiltonian
\begin{equation}
H={p_i^2\over 2M} +0.1E_J[1-\cos(2\pi x_i/a)]\; .
\end{equation}
That means, it is a delocalized quantum particle characterized by a
band structure $\epsilon(k)$. In the limit of small quantum
fluctuations of the phase, $E_C \lesssim 0.4 E_J$, the problem can be
treated in the tight-binding approximation\cite{Landau9}. This leads
to the following expression for the dispersion relation
\begin{equation}
\epsilon(k)=-{E_p\over 2}\cos(ka)\; ,
\label{dispersion}
\end{equation}
where the bandwidth is given by
\begin{equation}
E_p={8\over\pi}\sqrt{0.1E_JE_C}\exp(-2\sqrt{0.1E_J/E_C})\; ,
\label{tightbinding}
\end{equation}
and the effective mass of the vortices in the periodic potential by
\begin{equation}
m_{\text{eff}}^{-1}={a^2 E_p\over \hbar^2}\; .
\end{equation}
At stronger fluctuations the tight-binding approximation is
inadequate, and the bandwidth becomes of the order of $E_C$. We note
that $E_p$ can be interpreted as a transition amplitude. The exponent
of this amplitude can be also extracted from
Ref.~\onlinecite{LarkinOS}, where the rate of transitions between two
adjacent minima of the pinning potential was estimated. According to
Ref.~\onlinecite{LarkinOS}, this exponent is approximately
$2.25\sqrt{E_J/8E_C}$ which is about $10\%$ higher than the exponent
in Eq.~(\ref{tightbinding}).

In Eq.~(\ref{deviations}) we showed that each vortex moves in a
parabolic potential produced by the interaction with the other
vortices. The corresponding oscillation frequency of a particle having
effective mass $m_{\text{eff}}$ is
\begin{equation}
\omega_{\text{osc}}=\sqrt{{4\pi^2E_Jn\over W m_{\text{eff}}}} ={1\over
\hbar}\sqrt{2\pi^2E_JE_pna^2/W}\; ,
\end{equation}
and the mean-square oscillation amplitude is
\begin{equation}
\langle (x_i-x_i^0)^2\rangle={\hbar\over m_{\text{eff}}
\omega_{\text{osc}}}= a\sqrt{E_pW\over 8\pi^2E_Jn}\; .
\label{rmsamplitude}
\end{equation}

The quantum fluctuations implied by Eq.~(\ref{rmsamplitude}) lead to a
reduction of the effective pinning potential, which becomes now a
(periodic) function of $x_i^0$. Estimating the effective pinning, we
assume that the inter-vortex interaction on the scale of the lattice
constant $a$ is weak compared to $E_p$. In accordance with the
standard prescription of solid-state physics\cite{Landau9}, we
replace the quasi-wave vector in the dispersion relation
Eq.~(\ref{dispersion}) by an operator ${\hat p}/\hbar$, and consider
the Schr\"odinger equation for a quasiparticle with Hamiltonian
\begin{equation}
\tilde H=\epsilon\left(\frac{{\hat p}}{\hbar}\right) +{1\over 2}
m_{\text{eff}} \omega_{\text{osc}}^2 ({\hat x}-x_i^0)^2\; .
\label{quasiparticle}
\end{equation}
Here $\hat p$ and $\hat x$ are canonically conjugate variables. But
now we can view $\hat x$ as the momentum of some particle, moving in
the potential $\epsilon (p/\hbar)$ that is periodic in the coordinate
$p$ of the particle. Therefore, eigenstates of the Hamiltonian
(\ref{quasiparticle}) with various values of the ``quasi-momentum''
$x_i^0$ form bands. For each band, the energy is a periodic function
of $x_i^0$ with period $a$. At zero temperature, we are interested in
the lowest band, with the energy
\begin{equation}
U_p^{\text{eff}}(x_i^0)=U_p^{\text{eff}}\cos(2\pi x_i^0/a)\; .
\label{pinnpot}
\end{equation}
The value of $U_p^{\text{eff}}$ depends on the magnitude and form of
the periodic potential $\epsilon (p/\hbar)$. Using (\ref{dispersion})
and (\ref{tightbinding}), we find:
\begin{equation}
U_p^{\text{eff}}=E_J{a\over W}\sqrt{2nW\frac{E_p}{E_J}}
\exp\left(-\frac{2\sqrt 2}{\pi}{W\over a}\sqrt{{E_p\over E_J}{1\over
nW}}\right)\; .
\label{effectivepinning}
\end{equation}
Equation~(\ref{pinnpot}) gives the effective pinning potential for a
{\it single} vortex. It is worth noting that the pinning strength
diminishes with the increase of the equilibrium inter-vortex distance
$1/n$.

The approximations we employed in deriving the form
[Eq.~(\ref{pinnpot})] and amplitude [Eq.~(\ref{effectivepinning})] of
the pinning potential require a sufficiently wide band for the motion
of a vortex in the periodic potential. In other words, the exponential
factor in Eq.~(\ref{effectivepinning}) must be small. In the opposite
limit of negligible quantum fluctuations, the magnitude of the
effective potential is $0.2E_J$, and the function $U_p^{\text{eff}}
(x_i^0)$ has cusps at $x_i^0$ coinciding with the maxima of the bare
potential $U_p(x)$ defined in Eq.~(\ref{barepotential}). 
Each cusp in $U_p^{\text{eff}}(x_i^0)$ corresponds
to a jump of the coordinate of a {\it classical} vortex between the
minima of the potential $U_p(x)$.

One may get an idea of how effective the quantum smearing is, by
estimating $U_p^{\text{eff}}$ at $W/a=10$ and $E_C\simeq 0.4E_J$ which
is close to the limit of applicability of (\ref{tightbinding}).
Substitution of these values in Eq.~(\ref{effectivepinning}) yields
$U_p^{\text{eff}}\simeq 0.056\sqrt {nW}E_J\exp{(-3.58/\sqrt {nW})}$.
In this example, the effective pinning potential gets smaller than its
bare value at $1/n\gtrsim W/40$ which is always the case in practice.

At finite temperature $T\gtrsim\hbar\omega_{\text{osc}}$, we have to
consider the averaging of the periodic potential by quantum {\it and}
thermal fluctuations; these further reduce the pinning.
Summing the geometric series, we obtain
\begin{equation}
U_p^{\text{eff}}(T)=U_p^{\text{eff}} {1\over
1+\exp(-\hbar\omega_{\text{osc}}/T)}\; .
\end{equation}

We will now calculate the pinning of the rigid vortex chain. Each
vortex is subject to the potential Eq.~(\ref{pinnpot}). Summing over
the members of the chain [which are located at the positions shifted
by $\bar x$ from the equilibrium values (\ref{eqpositions})] leads to
\begin{eqnarray}
U_{\text{pin}}(\bar x)=\sum_{i=1}^N U_p^{\text{eff}}(x_i^0+\bar
x)\nonumber\\ =-U_p^{\text{eff}}(T)\cos(2\pi\bar x/a)
{\sin(N\pi/na)\over\sin(\pi/na)}\; .
\label{sinsum}
\end{eqnarray}

For commensurate values of the flux, i.e., if $1/na$ is integer, we get
\begin{equation}
U_{\text{pin}}(\bar x)=-U_p^{\text{eff}}(T)\cos(2\pi\bar x/a) N\; ;
\end{equation}
the pinning barrier is proportional to the total number of vortices,
i.e., the pinning is strong. In the immediate neighborhood of the
commensurate points, however, there are values of $n$ for which
\begin{equation}
n=\frac{\text{Int}(nL)}{a}\; .
\end{equation}
At these vortex densities, the numerator of Eq.~(\ref{sinsum})
vanishes, i.e., there is no pinning. The spacing between these zeroes
is approximately given by $na/L$, which may be less than
$1/L$. The rapid oscillations are caused by the fact that we are
considering a completely rigid vortex chain. If we neglect the
oscillations, and just look at the maxima of $U_{\text{pin}}$, it
turns out that the pinning strength behaves as $1/|n-n_0|$ close to
commensurate densities $n_0$.

\begin{figure}[h]
\begin{center}
\leavevmode \psfig{figure=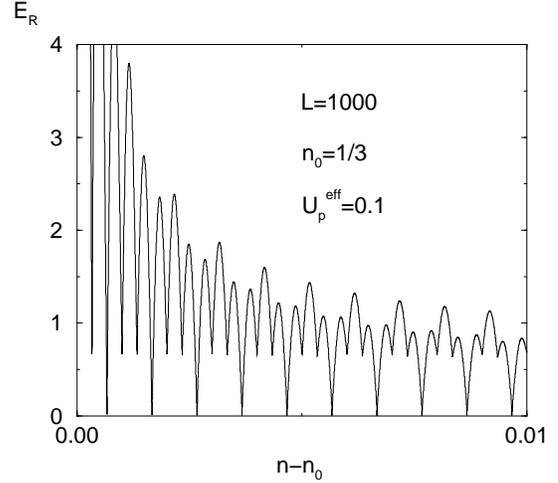,width=7cm}
\narrowtext
\caption{Activation energy entering the resistance as a function of
$n=BW/\Phi_0$. The energy plotted here is the sum of the boundary
pinning term Eq.~(\protect\ref{activation}), and twice the absolute
value of the bulk pinning term Eq.~(\protect\ref{sinsum}). Energy
units as in Fig.~\protect\ref{activationfigure}, and $n$, $n_0$ are 
measured in units of $1/a$.}
\label{activresis}
\end{center}
\end{figure}

The activation energy for the resistance can be estimated as the sum
of the amplitude of $U_{\text{pin}}(\bar x)$,
Eq.~(\ref{sinsum}), and the boundary pinning term $E_{\text{b}}(nL)$,
see Eq.~(\ref{activation}). The result is shown in
Fig.~\ref{activresis}.

Although each vortex is a quantum particle (as we have stressed at the
beginning of this section), the possibility of vortex permutations may
be safely neglected: for each vortex, $\langle (x_i-x_i^0)^2\rangle
\ll n^{-2}$, i.e., the oscillation amplitude is much less than the
inter-vortex distance.

\section{Compressible vortex chain}
\label{compressible}

In the last section we considered the case of the infinitely
long-range vortex-vortex interaction with the consequence that the
vortex chain was completely rigid. We will now discuss the importance
of screening and the resulting compressibility of the chain.

Screening of the vortex-vortex interaction in the Josephson array is
due to two effects: (1) screening by the magnetic field created by
currents flowing around the Josephson vortices (Meissner effect), and
(2) interaction with the vortex medium in the contact pads. To start
with, we consider the first of these two effects.

The distribution of currents flowing around a vortex depends on the
dimensionality of the system. The currents around a vortex line in a
three-dimensional superconductor drop off exponentially fast, the
characteristic length being the London penetration depth $\lambda_L$.
For a vortex in a superconducting film of small thickness
$s\ll\lambda_L$, the screening length\cite{Pearl} is
thickness-dependent, $\lambda=\lambda_L^2/s$. At a sufficiently large
distance from the center of a single vortex, $r\gtrsim\lambda$, the
spatial distribution of these currents is controlled by the Meissner
effect,
\begin{equation}
j_1(r)={\Phi_0c\over 4\pi^2 r^2}\; .
\label{Meissner}
\end{equation}
Note, that the distribution (\ref{Meissner}) is insensitive to the
short-scale structure of the two-dimensional vortex: it may be a
Josephson vortex in the array, as well as an Abrikosov vortex in the
contacts. In both cases, the currents induced by a vortex fall off as
$1/r^2$ at sufficiently large distances $r$ from its center, resulting
in a vortex-vortex interaction with a finite range
$\lambda_{\text{s}}$. In the specific case of a quasi-one-dimensional
Josephson junction array contacted by superconducting films, we may
estimate $\lambda_{\text{s}}$ by matching the current density $j_1$
with the density $j_2$ of the current flowing around a vortex in the
array,
\begin{equation}
j_2={2\pi^2E_Jc\over\Phi_0 W}\;.
\label{j2}
\end{equation}
The current density $j_2$ corresponds to a single-vortex contribution
to the phase gradient (\ref{gradient}). Equating
$j_1(\lambda_{\text{s}})\approx j_2$ leads us to the estimate
\begin{equation}
{\lambda_{\text{s}}\over W}\approx\sqrt{\Phi_0^2\over8\pi^4E_JW}\; .
\label{screening2}
\end{equation}
At distances $r\gtrsim\lambda_{\text{s}}$, the anisotropy of the
system is not important for the current distribution, and
Eq.~(\ref{Meissner}) is applicable. At smaller distances, the current
distribution is highly anisotropic. In the $x$-direction, the
screening length is $\lambda_{\text{s}}$, while in the $y$-direction
the currents are confined to the thin-film penetration depth
$\lambda$, which depends on the properties of the leads. Typically the
sheet superfluid density $\rho_s$ in the contacts exceeds greatly the
effective superfluid density $\rho_J$ in the Josephson junction array,
which leads to $\lambda\ll\lambda_{\text{s}}$.

The electrodynamic effect limiting the radius of interaction
considered above exists for any geometry of the contacts. The estimate
(\ref{screening2}) does not depend on the characteristics of the
contact material. However, the numerical coefficient in
(\ref{screening2}) is geometry-dependent. In Section \ref{experiment}
we will further refine the estimate (\ref{screening2}) for the
specific geometry of the experiment\cite{Oudenaarden1}.

In the above consideration, we have completely ignored the existence
of Abrikosov vortices in the contacts. This is acceptable only if
their currents do not overlap with the currents created by Josephson
vortices. The two current density fields are spatially separated if
the distance $d$ of the last row of the Abrikosov lattice to the edge
of the junction array (see Fig.~\ref{array}) exceeds $\lambda$. In the
opposite case $d\lesssim\lambda$, the Abrikosov lattice effectively
truncates the currents created in the contact by a Josephson
vortex. Indeed, a small shift of the lattice in the direction
perpendicular to the edge of the array is sufficient to compensate
these weak currents. To estimate the interaction potential range
$\lambda_{\text{s}}$ in this case, it is sufficient to deal with the
energy of the supercurrents,
\begin{equation}
E\simeq \frac{4\pi^2E_J\lambda_{\text{s}}}{W}+
\frac{\Phi_0^2d}{16\lambda_{\text{s}}\lambda}\; ,
\label{optimise}
\end{equation}
and neglect the magnetic fields the supercurrents create. In
Eq.~(\ref{optimise}), the first term corresponds to the energy of
currents in the array that flow in the region $|x|\lesssim
\lambda_{\text{s}}$ around the vortex. The second term is the energy
of the supercurrents in the contacts. These currents are truncated at
the position $d\simeq\sqrt{\Phi_0/B}$ of the first row of the
Abrikosov lattice, which numerically turns out\cite{Ternovsky} to be a
sound approximation.

Minimization of the energy Eq.~(\ref{optimise}) with regard to
$\lambda_{\text{s}}$ yields:
\begin{equation}
{\lambda_{\text{s}}\over W}\sim\sqrt{{\Phi_0^2\over 32\pi^2E_JW}
{d\over \lambda}}\; .
\label{screening1}
\end{equation}
This expression is valid for $d\lesssim\lambda$, and at $d\sim\lambda$
it reasonably well matches the estimate (\ref{screening2}). For
typical experimental values, Eqs.~(\ref{screening2}) and
(\ref{screening1}) yield a screening length for which $n^{-1} \ll
\lambda_{\text{s}} \lesssim L$. In other words, the vortex-vortex
interaction has long, but finite range, and the vortex chain is not
completely rigid.

We will now develop a continuum description of the compressible vortex
chain, i.e., we will express the energy of the chain in terms of the
deviations $u(x_i^0)=x_i-x_i^0$, and then go over to a deformation
field $u(x_i^0) \to u(x)$. The energy of the chain will be the sum of
a bulk pinning term and a boundary pinning term as before. In addition
to that, there will be an elastic energy term.

It is straightforward to express the bulk pinning term in terms of
$u(x)$:
\begin{eqnarray}
U_{\text{pin}}=-U_p^{\text{eff}}\sum_{i=1}^N\cos(2\pi
x_i/a)\nonumber\\ \approx
-nU_p^{\text{eff}}\int_{-L/2}^{L/2}dx\cos[{2\pi\over a}(u+\alpha x)]
\; ,
\label{U_pin}
\end{eqnarray}
where
\begin{equation}
\alpha=na\left[{1\over na}-\text{Int}\left({1\over
na}\right)\right]={n_0-n\over n_0}
\label{alpha}
\end{equation}
is a dimensionless measure of the deviation from the commensurate
value $1/n_0a=\text{Int}(1/na)$.

The ends of a long array ($L\gg \lambda_{\text{s}}$) act on the
compressible vortex chain as two independent sources of boundary
pinning. In the case of an almost rigid chain, $\lambda_{\text{s}}\gg
1/n$, we can find the pinning potential created by a single end (say,
the one corresponding to $x=L/2$) by a slight modification of
Eq.~(\ref{energy}). Namely, we introduce an exponential factor $\exp
[\gamma (x-L/2)]$ into the integrand, and replace the lower limit of
integration by $-\infty$. After that, we find the extremal value of
$\varphi_0$ as a function of the position of, say, the last vortex in
the chain $x_N$, and take the limit $\gamma\to +0$. This procedure
yields
\begin{equation}
U_{\text{b}}({\tilde u})=\frac{\pi^2}{6}\frac{E_J}{nW}
\left[4(n{\tilde u})^3-n{\tilde u}\right].
\label{endpin}
\end{equation}
Here for convenience we have introduced a new variable $\tilde u$
instead of the coordinate $x_N$,
\begin{equation}
{\tilde u} \equiv x_N-\left(\frac{L}{2}-\frac{1}{2n}\right), \nonumber
\end{equation}
and the coordinate $x_N$ is no more than one half-period away from
the end of the array, $|{\tilde u}|\leq 1/2n$. The top of the boundary
barrier is at $n{\tilde u}=-1/\sqrt{12}$, and its amplitude is
approximately $0.64E_J/nW$.

The elastic energy of vortices, which interact by long-range forces,
in the long-wavelength limit takes the form
\begin{equation}
U_{\text{el}}={K\over 2}\int_{-L/2}^{L/2}dx \left ({\partial u\over
\partial x}\right )^2\; .
\label{U_el}
\end{equation}
The elastic constant $K$ can be expressed through the vortex-vortex
interaction potential as
\begin{equation}
K=n^2\int_{-\infty}^{\infty}dx U_{v-v}(x)\;.
\label{elconstant}
\end{equation}
Note, that the potential $U_{v-v}(x)$ here is defined differently from
Eq.~(\ref{vortex-vortex}). Unlike in Eq.~(\ref{vortex-vortex}), we
remove the uncertainty in the definition of the potential by requiring
$U_{v-v}(x\to\pm\infty)=0$.

In order to estimate $K$, we adopt the following model for the
interaction potential:
\begin{equation}
U_{v-v}(x)=\frac{2\pi^2E_J}{W}\lambda_{\text{s}} \exp\left(
-\frac{|x|}{\lambda_{\text{s}}}\right),
\label{Umodel}
\end{equation}
which correctly reproduces the cusp [cf. Eq.~(\ref{vortex-vortex})] at
$|x|\ll \lambda_{\text{s}}$, and reaches zero at
$|x|\to\infty$. Within this model, we find
\begin{equation}
K\approx n^2\int_{-\infty}^{\infty}dx U_{v-v}(x)={4\pi^2E_J\over
W}(n\lambda_{\text{s}})^2\; .
\label{elconstantestimate}
\end{equation}
(The real interaction potential in the planar geometry of the contacts
considered above falls off as $x^{-2}$, rather than
exponentially. However, this should not significantly alter the
estimate). We complete the estimate of $K$ by adopting
Eq.~(\ref{screening2}) for the value of $\lambda_{\text{s}}$, which
yields
\begin{equation}
K\approx \frac{\Phi_0^2n^2}{2\pi^2}\; .
\label{elconstantestimate1}
\end{equation}
The elastic constant becomes softer, if the period of the Abrikosov
lattice in the contacts is smaller than $\lambda$, see
Eq.~(\ref{screening1}).

Varying $E=U_{\text{el}}+U_{\text{pin}}$ with respect to $u(x)$ leads
to the static sine-Gordon equation
\begin{equation}
K\left ({\partial^2 u\over \partial x^2}\right
)-nU_p^{\text{eff}}\sin\left({2\pi u\over a}-\alpha x\right)=0\; .
\label{sinegordon}
\end{equation}
This equation has been studied in many contexts, e.g.,
commensurate-discommensurate transitions in adsorbate
layers\cite{Chaikin}, here $\alpha$ is the difference of the lattice
constants of the substrate and the adsorbate. Another example is the
theory of long Josephson junctions \cite{Kulik} where $\alpha$ is
proportional to the magnetic field threading the junction.

\section{Phases}
\label{phases}

The behavior of a one-dimensional vortex chain is closely related to
that of an adsorbate layer\cite{Chaikin}: if the magnetic field is
commensurate, $n=n_0$, the vortex chain is commensurate with the
junction array. The activation energy of an elementary excitation at
$n=n_0$ is given by the energy to push one soliton into the
system. The length of such a soliton is given by
\begin{equation}
x_{\text{s}}={a\over 2\pi}\sqrt{K\over nU_p^{\text{eff}}}\; ,
\label{solitonlength}
\end{equation}
and its energy is
\begin{equation}
E_{\text{s}}={4a\over\pi}\sqrt{KnU_p^{\text{eff}}}\; .
\label{solitonenergy}
\end{equation}
A comparison of $x_{\text{s}}$ with the interaction radius
Eq.~(\ref{screening2}) yields $x_{\text{s}}/\lambda_{\text{s}}\simeq
(a/W)\sqrt {nW}\sqrt{0.4\pi^2E_J/U_p^{\text{eff}}}$; here we have
used the estimate (\ref{elconstantestimate1}) for the elastic
constant $K$. The applicability of Eq.~(\ref{sinegordon}) requires
$x_{\text{s}}\gtrsim\lambda_{\text{s}}$, and therefore
Eqs.~(\ref{solitonlength}), (\ref{solitonenergy}) are valid only if
the pinning potential is reduced by quantum fluctuations compared to
its classical value.

On moving $n$ away from $n_0$, the magnetic field tries to enforce a
period of the vortex lattice that is different from the period of the
pinning potential. The chain stays locked to a commensurate state up
to a critical value of $|n-n_0|$, or, in other terms, until $|\alpha|$
is less than some critical value $\alpha_C$. Below the threshold, at
$|n-n_0|\leq n_0|\alpha_C|$, the activation energy will diminish
linearly with increasing $|\alpha|$. This can be seen immediately from
the analogous situation in a long homogeneous Josephson junction where
quantized fluxons play the role of solitons. The energy to create the
first fluxon in the junction has some value at $H=0$, and decreases
linearly like $-M|H|$ with $|H|$; here $M>0$ is the magnetization of a
single fluxon, i.e., a constant. In a complete analogy with this, in
our case the energy to create a soliton in the commensurate phase is
of the form
\begin{equation}
E_{\text{s}}(n)=E_{\text{s}}\left[1-{|n-n_0|\over
n_0\alpha_C}\right]\; .
\label{cusp}
\end{equation}
Note that the energy $E_{\text{s}}(n)$ has a cusp-like dependence on
$n-n_0$. The point $n=n_0$ is special: the creation of a soliton or
anti-soliton costs the same energy. This situation is similar to the
line of particle-hole symmetry for a Mott insulator. Deviation from
the symmetry point makes creation of solitons, or anti-solitons
preferable. This violation of the symmetry is the origin of the
non-analytic dependence of $E_{\text{s}}$ on $n$. The critical values
of $\alpha$ at which the soliton energy turns zero are given by
\begin{eqnarray}
|\alpha|=\alpha_C\equiv {4\over\pi}\sqrt{nU_p^{\text{eff}}\over K}
\nonumber\\ ={8a\over x_{\text{s}}}\; .
\label{ncrit}
\end{eqnarray}

Above the threshold, at $|\alpha|>\alpha_C$, discommensurations will
exist: the chain is strained in the discommensurations, but this is
offset by the fact that the rest of the chain can stay in the minima
of the pinning potential. In this incommensurate phase, the
concentration of solitons is finite. Due to the solitons, the vortex
chain regains a finite compressibility $K_{\text{s}}$, which depends
on how far the system is tuned away from the critical points
$\alpha=\pm\alpha_C$. Without giving the details here, we note that
the dependence of the renormalized elastic constant on the control
parameter $\alpha$ can be presented in a parametric form\cite{Kulik},
as follows:
\begin{eqnarray}
{K_{\text{s}}\over K}&=&\frac{4}{\pi^2}
\frac{\frac{d}{d\gamma}\left[E(\gamma)/\gamma\right]}
{\frac{d}{d\gamma}\left[1/\gamma K(\gamma)\right]}\; , \nonumber\\
\frac{|\alpha|}{\alpha_C}&=&\frac{E(\gamma)}{\gamma}\; .
\label{Ks}
\end{eqnarray}
Here $K(\gamma)$ and $E(\gamma)$ are the complete elliptic integrals
of the first and second kind, respectively. The chain softens near the
critical points, where the proper expansion\cite{Chaikin} of
Eq.~(\ref{Ks}) yields:
\begin{equation}
{K_{\text{s}}\over K}
=\frac{8}{\pi^2}\frac{|\alpha|-\alpha_C}{\alpha_C}
\left[\ln\frac{\alpha_C}{|\alpha|-\alpha_C}\right]^2\; .
\label{Kscritical}
\end{equation}
The softening occurs, because the solitons in the chain are rare, and
the pair potential acting between them is exponentially small,
$U_{\text{s}}\sim E_{\text{s}}\exp(-x/x_{\text{s}})$. Far away form
the transition, at $|n-n_0|\gg \alpha_Cn_0$, the solitons overlap, and
$K_{\text{s}}= K$.

A finite voltage between the contacts to the array (see
Fig.~\ref{array}) is related, by the Josephson relation, to the
average velocity of vortices moving along the array. The transport of
a vortex through the system can be viewed as propagation of solitons
through the vortex chain. The availability of solitons in the chain
will clearly affect the resistance of the array. In the commensurate
phase, the soliton density is exponentially small at low temperatures,
$E_{\text{s}}(n)$ being the corresponding activation energy. In the
incommensurate phase, there are mobile solitons in the system even at
zero temperature. If one neglects the existence of boundary effects,
the activation energy $E_R$ of the observable quantity, viz.,
resistance, would coincide with the activation energy of a single
soliton. Hence, one would expect $E_R=E_{\text{s}}(n)$ in the
commensurate phase, and $E_R=0$ in the incommensurate phase.

It turns out, however, that boundary pinning modifies this
picture. First of all, it may affect the ground state of the vortex
chain. In the incommensurate phase, even weak boundary pinning will
lead to a deformation of a long compressible chain. In the
commensurate phase, the structure of the ground state starts to depend
on the ratio of the boundary pinning energy $E_{\text{b}}\sim E_J/nW$,
and the soliton energy $E_{\text{s}}(n)$; this ratio depends on the
bare parameters of the system, and may be small or large. Second, the
set of excited states the chain goes through during the elementary act
of a vortex transfer, also depends on the boundary pinning. These two
factors determine the dependence of $E_R$ on the characteristic
energies $E_{\text{s}}$ and $E_{\text{b}}$. We will analyze the
activation energy $E_R$ for both cases of small and large value of
this ratio.

\begin{figure}[h]
\begin{center}
\leavevmode \psfig{figure=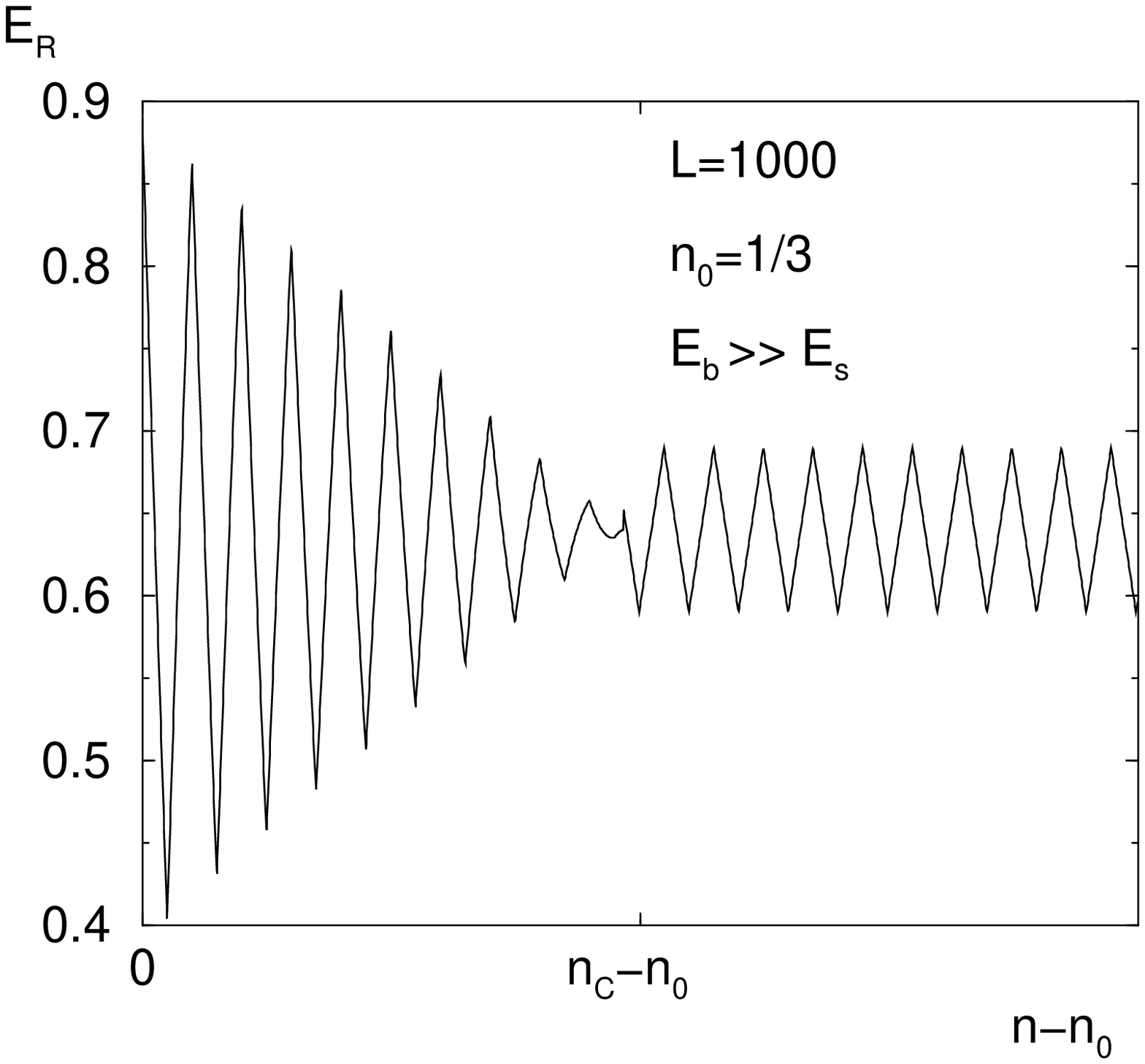,width=7cm}
\psfig{figure=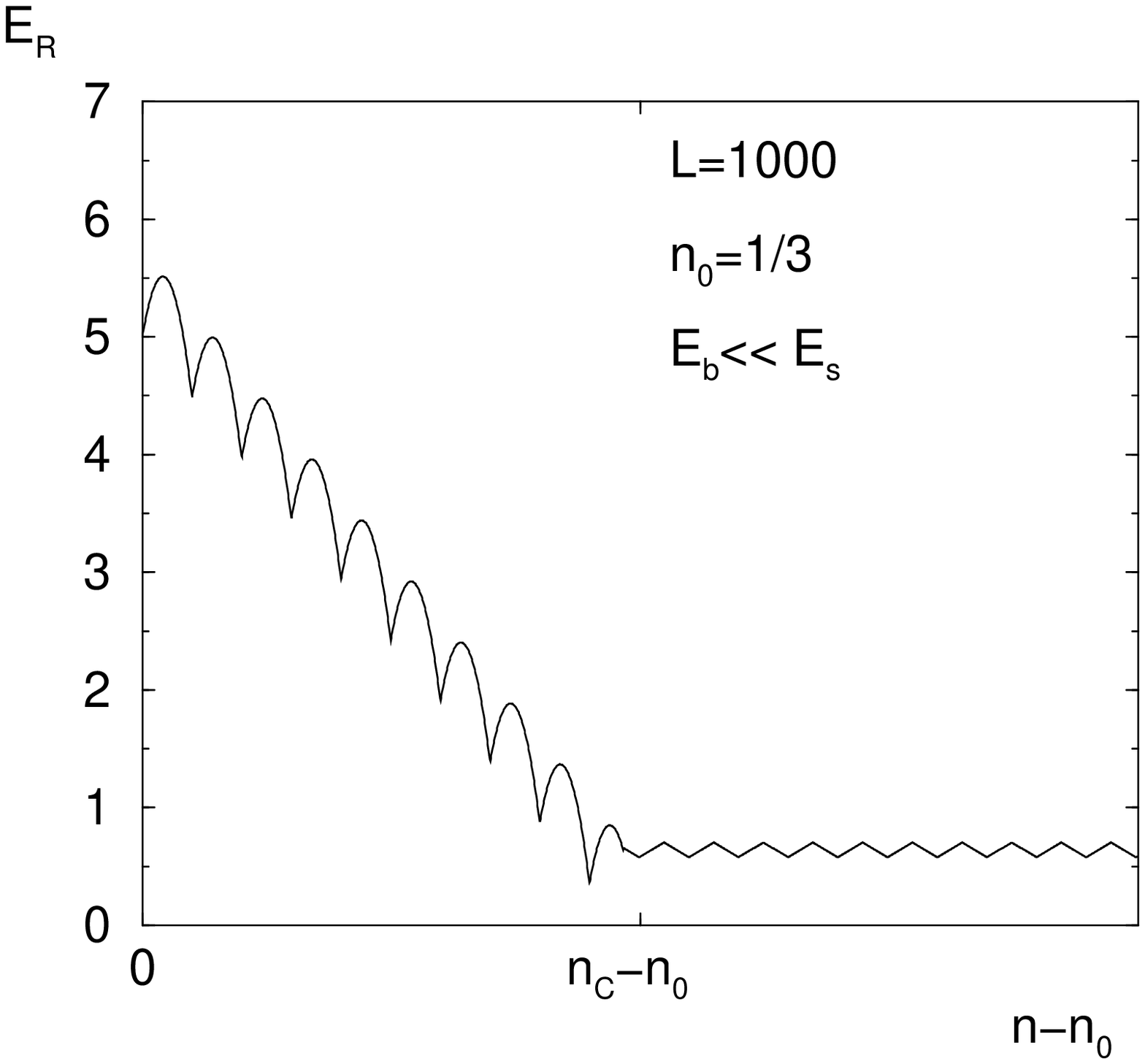,width=7cm}
\narrowtext
\caption{Phase diagram of the compressible vortex chain: activation
energy as a function of $n-n_C$. (a) $E_{\text{b}}\gg E_{\text{s}}$,
i.e., boundary pinning dominates ($E_{\text{s}}=0.5$). (b)
$E_{\text{b}}\ll E_{\text{s}}$, i.e., soliton formation energy
dominates ($E_{\text{s}}=5$). Energy units as in
Fig.~\protect\ref{activationfigure}, and the one-dimensional vortex densities
$n$, $n_0$, $n_C$ are measured in units of $1/a$.
On the incommensurate side of the 
transition, $n>n_C$, solitons will form spontaneously. The physics of the
incommensurate phase is therefore determined by boundary pinning and the
elastic energy and is the same for (a) and (b).}
\label{phasediagram}
\end{center}
\end{figure}

In the commensurate phase, and in the presence of strong boundary
pinning, $E_{\text{b}}>E_{\text{s}}(n)/2na$, the vortex chain in the
ground state will adjust itself to the length of the array to minimize
the pinning energy. This means there are solitons in the ground state,
unless $nL$ is an integer. The largest number of solitons in the
ground state occurs at a half-integer value of $nL$, and equals
$1/2na$. In this particular case, the chain without solitons is the
configuration with the highest energy that the system passes through
during a vortex transfer. In this state, the chain is not adjusted to
the boundary of the array, and the boundary pinning energy attains its
maximum value $E_{\text{b}}$. The difference of this energy from the
ground state is $E_R=E_{\text{b}}-E_{\text{s}}(n)/2na$. In the
opposite case of integer $nL$, there are $1/2na$ solitons in the
``saddle point'' state, and the activation energy reaches its maximum,
$E_R=E_{\text{b}}+E_{\text{s}}(n)/2na$, see Fig.~\ref{phasediagram}a.

If $E_{\text{b}}<E_{\text{s}}(n)$, there are no solitons in the ground
state of a commensurate chain. Moreover, during the process of a
vortex transfer through the array, there is at most one soliton in the
chain. Since a soliton changes the length of the vortex chain only by
$a$, the chain remains rigid on the scale of the inter-vortex distance
$1/n$. Therefore, we arrive at the following picture of the vortex
transfer. The passage of each soliton shifts the chain by $a$. The
transfer of a vortex requires the sequential passing of $1/na$
solitons. In this process, the chain moves as a rigid object in the
presence of boundary pinning. Thus, $E_R$ is the sum of
$E_{\text{s}}(n)$ and the boundary pinning energy
Eq.~(\ref{activation}) for a rigid chain,
\begin{equation}
E_R=E_{\text{s}}(n)+E_{\text{b}}(nL)\; ,
\label{ER}
\end{equation}
see Fig.~\ref{phasediagram}b. The soliton energy (\ref{cusp}) vanishes
at the boundaries of the commensurate phase. Before it vanishes, we
cross over to the case described in the previous paragraph.

We will now discuss the incommensurate phase, $n>n_C$. At the phase
transition, the soliton formation energy vanishes, and solitons will
start to form spontaneously. Correspondingly, the physics of the
incommensurate phase will be determined by boundary pinning and by the
elastic energy, and the behavior of the activation energy will be
identical for the two panels of Fig.~\ref{phasediagram}. The chain is
compressible; the elastic constant $K_{\text{s}}(n)$ is renormalized
down by solitons, see Eq.~(\ref{Ks}). The adjustment of the vortex
chain to the length of the array leads to a finite deformation. The
corresponding elastic energy can be found with the help of
Eq.~(\ref{U_el}) with $K$ replaced by $K_{\text{s}}(n)$. The maximum
value of the deformation $\partial u/\partial x =1/2nL$ corresponds to
half-integer $nL$, and the elastic energy associated with it is
$K_{\text{s}}(n)/8n^2L$. For large $L$, this energy is inevitably
smaller than the boundary pinning potential (\ref{endpin}). To
initiate an elementary act of vortex transport through the array, a
shift of the end vortex through the maximum of the potential
(\ref{endpin}) should occur. This varies the deformation of the chain
by $1/n\sqrt 3$, or by $(1-1/\sqrt 3)/n$. The corresponding elastic
energy in both cases is the same, and is equal to
$K_{\text{s}}(n)(1-2/\sqrt 3)^2/8n^2L$. The net variation of the
elastic energy involved in the described shift of the vortex, is
\begin{equation}
\delta U_{\text{el}}=\frac{K_{\text{s}}(n)}{8n^2L}
\left[\left(1-\frac{2}{\sqrt 3}\right)^2 -1\right]\approx
-0.12\frac{K_{\text{s}}(n)}{n^2L}\; .
\label{deltaU}
\end{equation}
To obtain the activation energy for the resistance $E_R$ at this
particular value of $nL$, one should add $\delta U_{\text{el}}$ to the
boundary pinning amplitude. At some other values of $nL$, the
variation in the elastic energy involved in the process of passing the
boundary barrier, attains its maximum value $-\delta U_{\text{el}}$.
Thus, the resistance activation energy oscillates between two values,
\begin{equation}
E_R\approx 0.64\frac{E_J}{nW}\pm 0.12\frac{K_{\text{s}}(n)}{n^2L}
\label{elasticandboundary}
\end{equation}
with the period $\Delta n = 1/L$.

\section{Comparison with the experiment}
\label{experiment}

The resistance $R$ of a number of arrays of Josephson junctions was
measured in the geometry depicted in Fig.~\ref{array} in the presence
of a magnetic field. Arrays with lengths $L$ varying between $100a$
and $1000a$, and widths $W$ of $7a$ and $3a$ were studied. The
characteristic Josephson energy for all the samples was about $1$K,
with the ratio $E_J/E_C$ varying within the limits
$0.7$ to $2.8$. (The details of sample preparation as well as the
experimental techniques can be found in 
Ref.~\onlinecite{Oudenaarden1}). The main qualitative feature of the
field dependence of $R$ consists in the existence of a finite region
of magnetic flux densities $n$ around the commensurate value
$n_0=1/3a$, where the resistance is strongly suppressed (Mott phase
for the system of quantum vortices). The width of this region becomes
smaller with the increase of the ``quantum parameter'' $E_C/E_J$,
see Ref.~\onlinecite{Oudenaarden1}, in agreement with the notion of the Mott
transition.

\begin{figure}[h]
\begin{center}
\leavevmode \psfig{figure=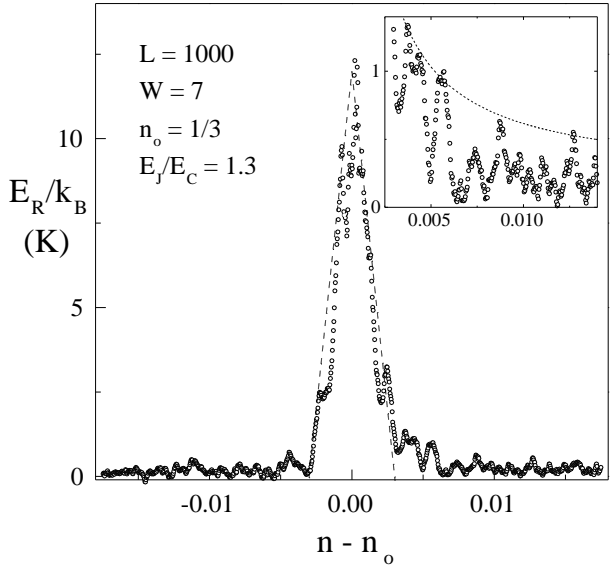,width=8cm}
\narrowtext
\caption{Activation energy of the resistance of an array consisting of
$1000\times 7$ cells with parameters $E_J=0.9$K and $E_C=0.7$K. The
one-dimensional vortex densities $n$, $n_0$ are measured in units of
$1/a$. The dashed line is a fit to the data, to extract the width of the
Mott region. The cusp-like part of the figure
corresponds to the Mott phase. Additional wiggles on that part may be
related to boundary effects, see Section~\protect\ref{phases} and
Fig.~\protect\ref{phasediagram}b. Inset: activation energy outside the
Mott phase. The zeros of $E_R(n)$ at $n-n_0>0.006$ indicate the
restored rigidity of the vortex chain. Note that in qualitative
agreement with Fig.~\ref{activresis}, the maxima of $E_R(n)$ decrease with
increasing $|n-n_0|$; the dotted line in the inset is a guide to the eye.}
\label{data}
\end{center}
\end{figure}

Within the Mott phase, the resistance clearly displays an activated
behavior, with the activation energy $E_R$ strongly depending on the
deviation $|n-n_0|$ from the point of exact commensurability. In
Fig.~\ref{data} we present new data for the activation energy for our
longest sample, $L=1000a$, with parameters $W=7a$, $E_C=0.7$K, and
$E_J=0.9$K. For each value of $n$, the activation energy $E_R$ was
determined from the measured temperature dependence of the array
resistance. The measurement was performed in the linear regime, at a
small transport current. For this sample, the commensurate phase
around the point $n_0=1/3a$ exists in the domain
$|\alpha|<\alpha_C\approx 0.009$. The maximal value of the activation
energy, $E_R\approx 12$K, is reached at the commensurability
point. Outside the Mott phase region, the resistance exhibits strong
oscillations; the activation energy vanishes almost periodically, with
the period $\Delta n = 1/L$. We find two aspects of this data
striking.

Firstly, the regions of $n$ corresponding to the Mott phase are
extremely narrow ($\alpha_C\simeq 10^{-2}$). In the conventional
picture, this would imply a weak interaction between the particles
(compared to the one-particle band structure energies). Consequently,
within the Mott phase the activation energies for particle transport
must be also small. Quite contrary, the observed value of the
resistance activation energy is about one order of magnitude larger
than the energies $E_C$ and $E_J$, which determine the single-vortex
band spectrum.

Secondly, the resistance $R(n)$ exhibits strong oscillations with the
period $\Delta n = 1/L$ outside the Mott region. These oscillations
would not be expected in a model of almost-free quasiparticles within
the delocalized phase.

These two observations find a natural explanation in our model, which
explicitly accounts for the long-range interaction forces between the
vortices.

In order to perform a detailed quantitative comparison of the theory
and experiment, first we improve the estimates (\ref{screening2}) and
(\ref{elconstantestimate1}) for the interaction range $\lambda_{\text
s}$ and the elastic constant $K$ respectively. In the
experiment\cite{Oudenaarden1}, the contact bars were made by shorting
the junctions at the lower and upper border of the array. We therefore
model the bars by superconducting strips of average width 
$a\approx 10^{-4}$cm (the lattice constant of the array). The condition
$\lambda_{\text{s}}\gg W$ allows us to neglect non-local effects
in the solution of the magnetostatic problem\cite{Orlando}, and to
express $\lambda_{\text s}$ in terms of the self-inductance ${\cal L}$
of the two-wire system. In addition, as the distance $W$ between the
wires exceeds significantly their width $a$, we can use the
textbook\cite{Jackson} formula ${\cal L} =4\ln (W/a)$. As a result,
Eq.~(\ref{screening2}) is replaced by
\begin{equation} 
\lambda_{\text{s}}
=W\sqrt{\Phi_0^2\over 16\pi^2E_JW\ln(W/a)}\simeq 270a\; . 
\label{lambdaestimate} 
\end{equation} 
(The resulting numerical value here refers to the parameters of the sample
of Fig.~\ref{data}).
In the same approximation, the model form of the potential
(\ref{Umodel}) becomes exact. With the help of
Eq.~(\ref{lambdaestimate}), the estimate (\ref{elconstantestimate1})
of the elastic constant can also be refined. For convenience, we give
here the value of the product $aK$, instead of the value of the
elastic constant:
\begin{equation} 
aK\approx \frac{\Phi_0^2n^2a}{4\ln
(W/a)}\simeq3.5\times 10^4 {\text K}\; . 
\label{Kestimate} 
\end{equation}

The theoretical results (\ref{lambdaestimate}) and (\ref{Kestimate}) do
not have any adjustable parameters, and are obtained within controllable
approximations. In contrast with this, Eq.~(\ref{effectivepinning}) for
the effective pinning cannot be used for the quantitative comparison with
the experiment\cite{Oudenaarden1}: The tight-binding approximation we have
used in Section \ref{effpinning} to estimate the suppression of
the pinning potential is not applicable to the case $E_C\approx E_J$.
(We note that Eq.~(\ref{pinnpot}) still allows one to reproduce the
correct trend in the variation of $\alpha_C$ with the ratio
$E_C/E_J$). Therefore, we proceed in the following way. First, we find
the soliton length $x_{\text s}$ from the experimental values of
$\alpha_C$. Then, using the theoretical value of $aK$
and the value of $x_{\text s}$ extracted from the data, we find the
renormalized pinning potential $U_p^{\rm eff}$ and the soliton
activation energy $E_{\text s}$. We will check that the renormalized
pinning potential is indeed substantially lower than its bare value
$0.1E_J$. Finally, we will relate the found value of $E_{\text s}$ to
the activation energy of the resistance for the experimental sample.

The values of $\alpha_C$ and $a\approx 10^{-4}$cm found experimentally
allow us to estimate the soliton length from Eq.~(\ref{ncrit}):
\begin{equation}
x_{\text{s}}=\frac{8a}{\alpha_C}= 890a \approx 0.09 {\text{cm}}\; .
\label{xs}
\end{equation}
This length is really large. In fact, $x_{\text{s}}$ is about three
times the length of the array $L=300a$ used in
Ref.~\onlinecite{Oudenaarden1} to extract the activation energy
$E_R$. This may explain why the values of $E_R$ found there are
systematically lower than the activation energy for the longest array,
see Fig.~\ref{data}. A single soliton consists of about $300$ vortices,
and therefore its activation energy may exceed easily the
single-vortex energy scales. Note also that $x_{\text{s}}$ exceeds
considerably the interaction radius (\ref{lambdaestimate}) which gives
us confidence in the applicability of the sine-Gordon equation
(\ref{sinegordon}). The effective pinning potential, according to
Eq.~(\ref{solitonlength}), can be found as:
\begin{equation}
U_p^{\rm eff}=\left(\frac{a}{2\pi x_{\text s}}\right)^2\frac{aK}{na}
\simeq 3.4\times 10^{-3} {\text K}\; .
\label{Uestimate}
\end{equation}
This energy is at least one order of magnitude smaller than its bare
value $0.1E_J$, see Eq.~(\ref{barepotential}). The reduction is
apparently due to the quantum zero-point motion of individual
vortices. Again, the strong suppression of the pinning energy
guarantees the harmonic form of the pinning potential (\ref{pinnpot}),
and hence allows us to use the sine-Gordon equation for the
solitons. Finally, using Eqs.~(\ref{solitonlength}) and
(\ref{solitonenergy}), we find the soliton energy at the
commensurability point:
\begin{equation}
E_{\text s}=\frac{\alpha_C}{4\pi^2}aK\simeq 8\; {\text K}\; .
\label{Esestimate}
\end{equation}
This energy exceeds significantly the boundary pinning
energy. According to Eqs.~(\ref{activation}) and (\ref{ER}), the
latter contributes to $E_R$ less than $0.5$K. We neglect this
contribution, and therefore identify $E_R$ with the energy
$E_{\text{s}}$ of the formation of a soliton. The calculated value
(\ref{Esestimate}) is somewhat lower than the measured $E_R$. Still,
we find the agreement quite impressive, having in mind the huge value
of the elastic constant (\ref{Kestimate}), calculated without any
adjustable parameters.

The large value of the elastic constant (\ref{Kestimate}) results from the
long-range nature of the inter-vortex interaction forces. In fact, the
vortex chain in the incommensurate phase is so rigid, that
Eq.~(\ref{elasticandboundary}) is inapplicable in the case of a sample
only a thousand cells long. Away from the transition point 
($|\alpha|=\alpha_C$), the
elastic term in Eq.~(\ref{elasticandboundary}), which is supposed to be a
small correction, is about $70$ times larger than the ``main'' boundary
pinning term. Therefore, for the conditions of the experiment, the
incommensurate phase is well described by the model of a rigid vortex
chain, see Section \ref{rigid}. This immediately explains the strong
oscillations of the activation energy with the period $\Delta n= 1/L$ in
the incommensurate phase, see Eqs.~(\ref{activation}) and (\ref{sinsum}).

There is a clear resemblance between the experimentally measured curve
of $E_R(n)$ (inset in Fig.~\ref{data}), and the curve in
Fig.~\ref{activresis}, simulated with the help of
Eqs.~(\ref{activation}) and (\ref{sinsum}). In agreement with the
model of a rigid chain, the minima of $E_R(n)$ reach zero at
$|n-n_0|\gtrsim 0.005$, and the maxima of $E_R(n)$ decrease with
increasing $|n-n_0|$. The boundary term Eq.~(\ref{activation}) has a
maximum of $\approx 0.5$K and accounts for the main contribution to
the maxima of $E_R$ at $n-n_0>0.006$. The vortex chain softens up
only in a very narrow region around the transition point, so that the
crossover region is of the order of $\Delta n$, see the inset in
Fig.~\ref{data}.

To end this section, we reiterate that in the experiment the
commensurability point $n_0=1/3a$ was reached in the sample with
$W=7a$, which means the chain is stable against the formation of a
zigzag structure at $1/nW=0.43$. According to Eq.~(\ref{zigzag}), for
a continuous system, the zigzag instability would already occur at
$1/nW=0.65$, i.e., before the density $n_0=1/3a$ is reached. Since the
experimental data show no indication of a qualitative difference
between the arrays with $W=3a$ and $W=7a$, we conclude that the array
width $W=7a$ is narrow enough to allow suppression of the instability
by the effects of discreteness.

\section{Discussion}
\label{discussion}

An external magnetic field applied to an array of Josephson junctions
allows one to introduce vortices into it. A sufficiently weak field creates
a linear chain of vortices in the quasi one-dimensional array. The ratio
between the periods of the vortex chain and the array of
Josephson junctions is controlled by the value of the magnetic field. The
commensurate phase corresponds to the vortex analogue of a Mott insulator. 
Within this phase, the
elementary excitation is a soliton consisting of a number of individual
vortices. The finite gap energy for the soliton translates into a finite
activation energy of the resistance of the array. Each soliton transfers a
fraction of the flux quantum through the array. In the incommensurate
phase, the spontaneous proliferation of solitons and anti-solitons leads
to the formation of a one-dimensional vortex liquid. This results in a
finite vortex-flow resistance of the array. 

In this paper we have analyzed the commensurate-incommensurate
transition for a one-dimensional vortex system in detail. The size and
energy of the vortex solitons, which drive the transition, depend on
two parameters of the vortex chain. These parameters are: the elastic
constant, and the pinning potential existing due to the discreteness
of the array of Josephson junctions. The long-range nature of the
vortex-vortex interaction leads to a large value of the elastic
constant. On the other hand, the zero-point motion of each quantum
vortex leads to a considerable suppression of the pinning
potential. As a result, the size of the solitons turns out to be
extremely large, about $300$ vortices under the conditions of the
experiments reported in this paper and in
Ref.~\onlinecite{Oudenaarden1}. This enables us to treat the
transition in the framework of the classical theory\cite{Chaikin}. Our
theory explains quantitatively the main experimental observations.

We would like to conclude with the following remark: a quasi
one-dimensional array of small superconducting islands connected by
Josephson junctions can be used to study quantum phase transitions in
two complementary ways. The first way relies on the control of the
charge state of the islands by an external gate. In this case, a
transition between the charge-localized and charge-delocalized phases
can be observed in principle. The localized phase is a Mott insulator,
with a finite gap for charge solitons, which play the role of
elementary excitations. The delocalized phase behaves as a
one-dimensional Luttinger liquid (see, e.g., Ref.~\onlinecite{Larkin}
and references therein). The experimental observation of the two
phases and the transition is difficult, as it is virtually impossible
to avoid the existence of random offset charges, which introduce
strong disorder into the system. The other way is to study the
commensurate-incommensurate transition in a system of vortices induced
in the array by an external magnetic field (the case studied in this
paper). This transition belongs to the same universality class as the
Mott transition for charge delocalization. A great advantage of the
vortex system is that it is virtually disorder-free. However, due to
the large size of the solitons driving the transition, the critical
region around the phase transition point is extremely narrow for the
arrays studied experimentally. To widen the critical region, one
should find a way to reduce the vortex-vortex interaction
strength. That would open new possibilities of experimental
investigations of the Luttinger liquid which is formed on the
incommensurate side of the transition. The properties of the liquid
are expected to depend crucially\cite{Fisher} on the value of the
fractional flux carried by the solitons.

\acknowledgements

We would like to thank R. Fazio, K.~K. Likharev, Yu. Makhlin,
Yu.~V. Nazarov, and T. Orlando for discussions. C.B. would like to
thank the Theoretical Physics Institute of the University of Minnesota
for its hospitality. L.G. acknowledges hospitality of Delft Technical
University where part of this work was performed. The work at the
University of Minnesota was supported by NSF Grants DMR-9731756 and
DMR-9812340.

\end{multicols}

\begin{thebibliography}{99}

\bibitem[*]{Basel} Address after Oct. 1: Department of Physics and 
Astronomy, University of Basel, Klingelbergstrasse 82, CH-4056 Basel,
Switzerland.

\bibitem{Simanek94} E. Simanek, {\it Inhomogeneous Superconductors}
(Oxford University Press, New York, 1994).

\bibitem{FazioS} R. Fazio and G. Sch\"on, Phys. Rev. B {\bf 43}, 5307 (1991).

\bibitem{Geerligs} L.~J. Geerligs, M. Peters, L.~E.~M. de Groot,
A. Verbruggen, and J.~E. Mooij, Phys. Rev. Lett. {\bf 63}, 326 (1989).

\bibitem{LarkinOS} A.~I. Larkin, Yu.~N. Ovchinnikov, and A. Schmid,
Physica B {\bf 152}, 266 (1988).

\bibitem{EckernS} U. Eckern and A. Schmid, Phys. Rev. B
{\bf 39}, 6441 (1989).

\bibitem{Elion} H.~S.~J. van der Zant, F.~C. Fritschy, W.~E. Elion, L.~J.
Geerligs, and J.~E. Mooij, Phys. Rev. Lett. {\bf 69}, 2971 (1992).

\bibitem{Delftconf} For a number of articles see the Proceedings of the NATO
Advanced Research Workshop on {\it Coherence in Superconducting Networks},
edited by J.~E. Mooij and G. Sch\"on, Physica B {\bf 152} (1988).

\bibitem{Lobb} C.~J. Lobb, D.~W. Abraham, and M. Tinkham, Phys. Rev. B
{\bf 27}, 150 (1983).

\bibitem{Oudenaarden1} A. van Oudenaarden and J.~E. Mooij,
Phys. Rev. Lett. {\bf 76}, 4947 (1996); A. van Oudenaarden, B. van
Leeuwen, M.~P.~M. Robbens, and J.~E. Mooij, Phys. Rev. B {\bf 57}, 11684
(1998).

\bibitem{FisherFisher} M.~P.~A. Fisher, B.~P. Weichman, G. Grinstein, and
D.~S. Fisher, Phys. Rev. B {\bf 40}, 546 (1989).

\bibitem{Kardar} M. Kardar, Phys. Rev. B {\bf 33}, 3125 (1986).

\bibitem{critical} The conditions under which the
superconductor-insulator transition occurs were considered in a number
of papers that study a more general model, which includes the
capacitance $C_g$ of the grains to a gate. Probably the most reliable
estimate of the critical value of $E_C/E_J$ for the limiting case
$C_g\to 0$ considered here is provided by direct Monte Carlo
simulation, see J.~V. Jose and C. Rojas, Physica B {\bf 203}, 481 (1994).
The simulation yields $E_C/E_J\approx 1.4$.

\bibitem{DeGennes} P.~G. De Gennes, {\it Superconductivity of Metals and
Alloys} (W.~A. Benjamin, New York, 1966).

\bibitem{footnote} Taking into account the vortex core energies leads
to a small field-independent shift in the argument of the function
$\text{Int}(x)$. We neglect this shift here and in the rest of the paper. 

\bibitem{Likharev} K.~K. Likharev, Zh. Eksp. Teor. Fiz. {\bf 61}, 1700
(1972) [Sov. Phys. JETP {\bf 34}, 906 (1972)].

\bibitem{Landau9} L.~D. Landau and E.~M. Lifshitz, {\it Statistical Physics:
Part II}, p. 230f. (Pergamon Press, Oxford, 1980).

\bibitem{Pearl} J. Pearl, Appl. Phys. Lett. {\bf 5}, 65 (1964).

\bibitem{Ternovsky} F.~F. Ternovskii and L.~N. Shekhata, Zh. Eksp. Teor. Fiz.
{\bf 62}, 2297 (1972) [Sov. Phys. JETP {\bf 35}, 1202 (1972)].

\bibitem{Chaikin} P.~M. Chaikin and T.~C. Lubensky, {\it Principles of
Condensed Matter Physics} (Cambridge University Press, Cambridge, 1995).

\bibitem{Kulik} I.~O. Kulik and I.~K. Yanson, {\it The Josephson Effect in
Superconducting Tunneling Structures} (Israel Program for
Scientific Translations, Jerusalem, 1972).

\bibitem{Orlando} R.~D. Bock, J.~R. Phillips, H.~S.~J. van der Zant, and 
T.~P. Orlando, Phys. Rev. B {\bf 49}, 10009 (1994).

\bibitem{Jackson} J.~D. Jackson, {\it Classical Electrodynamics}, p. 262
(Wiley, New York, 1975).

\bibitem{Larkin} L.~I. Glazman and A.~I. Larkin,
Phys. Rev. Lett. {\bf 79}, 3736 (1997). 

\bibitem{Fisher}
M.~P.~A. Fisher and L.~I. Glazman, {\it Transport in a one-dimensional
Luttinger liquid}, in: {\it Mesoscopic Electron Transport}, edited by
L.~L. Sohn, L.~P. Kouwenhoven, and G. Sch\"on, p. 331 (Kluwer, The
Netherlands, 1997).

\end{thebibliography}
\end{document}